\documentclass[prl,twocolumn,showpacs,amsmath,amssymb]{revtex4}
\usepackage{graphicx}
\usepackage{dcolumn}
\usepackage{bm}
\begin{document}
\title{Microscopic Model of Critical Current Noise in Josephson Junctions} 
\author{Magdalena Constantin and Clare C. Yu}
\affiliation{
$^1$ Department of Physics and Astronomy, University of California, 
Irvine, California 92697-4575}
\date{\today}
\begin{abstract}
We present a simple microscopic model to show how fluctuating 
two--level systems in a Josephson junction tunnel barrier of thickness 
$L$ can modify the potential energy of the barrier and produce critical 
current noise spectra. We find low frequency $1/f$ noise that goes as $L^5$. 
Our values are in good agreement with recent experimental measurements 
of critical current noise in Al/AlO$_{x}$/Al Josephson junctions. 
We also investigate the sensitivity of the noise on the nonuniformity 
of the tunnel barrier.
\end{abstract}

\maketitle
Remarkable progress has been achieved in making
high--quality Josephson junction qubits
\cite{Vion2002,Yu2002,Chiorescu2003,Pashkin2003,Berkley2003,McDermott2005},
though sources of noise and decoherence continue to be problematic.
Recent experiments \cite{Simmonds2004,Martinis2005} indicate that a dominant
source of decoherence is two--level systems (TLS) in the insulating
barrier of the tunnel junction as well as in the dielectric material
used to fabricate the circuit. These TLS fluctuators produce low 
frequency $1/f$ critical current noise $S_{I_c}$ 
\cite{Wellstood,Savo1987,Wellstood2004,Harlingen2004,Mooij2006}.
However, a simple microscopic model showing this is still missing. 
In a common scenario for $S_{I_c}$ 
\cite{Harlingen2004}, defects in the oxide tunnel barrier affect nm 
sized conducting channels. But how can defects a few angstroms in width 
have a noticeable effect on a superconducting wave function with a
micron sized coherence length that is orders of magnitude larger than the
perturbing defect? The answer is that the tunneling current is
exponentially sensitive to perturbations of the tunnel barrier. 
We have confirmed this with a microscopic calculation 
of $S_{I_c}$ due to fluctuating TLS in the barrier
and obtain good quantitative agreement with experiment.

Previous theoretical work postulated that the qubit was coupled to 
fluctuating defects by putting a coupling term into the Hamiltonian 
\cite{Simmonds2004,Ku2005,Shnirman2005,Faoro2005,Faoro2006,Faoro2006b}, 
but no one has shown how this coupling arises microscopically.
In this Letter we calculate the $1/f$ critical current noise $S_{I_c}$ 
due to thermally fluctuating TLS that have electric dipole moments. 
We assume that the current $I$ through the Josephson junction is given by
$I=I_c\sin\delta$ where $I_c$ is the critical current and
$\delta$ is the phase difference between the superconductors. In a qubit the
junction is small ($\lambda_J\gg\sqrt{A}$ where $\lambda_J$ is the Josephson
penetration depth and $A$ is the area of the junction) so that the phase
difference is uniform in the plane of the junction.  We start by
calculating how a dipole modifies the junction's potential barrier 
$U({\bf r})$. We then use a WKB formalism to compute the tunneling matrix
element ${\cal T}_{LR}\sim\exp\left(-\sqrt{U}\right)$ 
between the left (L) and right (R) electrodes. 
The critical current $I_c$ is proportional to 
$\langle|{\cal T}_{LR}|^2\rangle$ averaged over the junction \cite{AB1963}. 
We consider elastic electron tunneling where different orientations of the 
dipole correspond to different values of ${\cal T}_{LR}$ and hence, $I_c$.
We can obtain $S_{I_c}$ since each fluctuating dipole behaves as a random 
telegraph variable that has a Lorentzian noise spectrum. By averaging over 
the standard TLS distribution, and each dipole's orientation and position 
along the $z-$axis, we obtain $S_{I_c}$. At low frequencies we find $1/f$ 
behavior for $S_{I_c}$, and our values are in good agreement with the 
corresponding experimental values \cite{Martinis1992,Zorin1996,Mooij2006}. 
Our model predicts that the noise is very sensitive to the tunnel 
barrier thickness $L$ and that $S_{I_c}\sim L^5$, implying that the 
noise can be greatly reduced by decreasing $L$.

In our model the TLS sits in the insulating tunnel barrier with
an electric dipole moment ${\bf p}$ consisting
of a pair of opposite charges separated by a distance $d$. 
The superconducting electrodes are located at $z=0$ and $z=L$ 
and kept at the same potential. The angle between ${\bf p}$ and 
the tunneling direction ($z$--axis) is $\theta_0$. The TLS
has a double--well potential with a Hamiltonian \cite{Phillips} 
$H_0=\frac{1}{2}(\Delta \sigma_z+\Delta_0 \sigma_x)$
where $\Delta_0$ is a tunneling matrix element, 
$\Delta$ is the energy difference between the right and left wells,
and $\sigma_{x,z}$ are the Pauli spin matrices. The energy eigenvalues
are $\pm E/2$ where $E=\sqrt{\Delta^2 + \Delta_0^2}$. 
The TLS couples to the strain field. 
So an excited two-level system can decay to the ground state by 
emitting a phonon. The longitudinal relaxation rate is given by \cite{Phillips} 
$T_1^{-1}=a E\Delta_0^2\coth(E/2k_B T)$, where the prefactor $a$ 
is a material dependent constant. The distribution of TLS parameters 
can be expressed in terms of $E$ and $T_1$:
$P(E,T_1)=P_0/(2T_1\sqrt{1-\tau_{min}(E)/T_1})$ 
\cite{Hunklinger1986,Phillips} where $P_0$ is the TLS density of states. 
The minimum relaxation time $\tau_{min}(E)=T_1(E=\Delta_0)$ corresponds to a symmetric 
double--well potential. 

We start by calculating the junction barrier potential energy $U$ when
it is distorted by the presence of TLS with electric dipole moments 
\cite{Schmidlin1966,Scalapino1967}. We use cylindrical coordinates 
($\rho,\phi,z$) with ${\bf \hat{z}}$ normal to the plane of the junction. 
For a square barrier, $U(\rho,\phi,z)=U_0-eV_{dip}(\rho,\phi,z)$, 
where $V_{dip}(\rho,\phi,z)$ is the contribution of the electric dipole 
to the barrier potential and $U_0$ is the height of the unperturbed 
uniform square barrier. We use Green's functions \cite{Jackson} to calculate 
$V_{dip}(\bm{r})=\int{G({\bf r,r^{\prime}})\alpha({\bf r^{\prime}})
\mbox{d}^3{\bf r^{\prime}}}$, where the charge density 
$\alpha({\bf r})=\sum_{i=1,2}{q_i \delta^3({\bf r-r_i})}$ with the
positive charge $q_1 = |{\bf p}|/d$ at ${\bf r_1}=z_0{\bf \hat{z}}$,
and the negative charge $q_2=-q_1$ at 
${\bf r_2}=d\;\mbox{sin}\theta_0{\bf\hat{x}} +
(z_0+d\mbox{cos}\theta_0){\bf\hat{z}}$.
The Green's function is \cite{Jackson} 
$G({\bf r},{\bf r^{\prime}})=1/(\pi\epsilon L)\times
\sum_{n=1}^{\infty}\sum_{m=-\infty}^{\infty}\sin(n\pi z/L)\sin(n\pi z^{\prime}/L)
\exp[im(\phi-\phi^{\prime})]\times I_m(n\pi\rho_{<}/L)K_m(n\pi \rho_{>}/L)$,
where $\rho_{<}=\min(\rho,\rho^{\prime})$, $\rho_{>}=\max(\rho,\rho^{\prime})$, 
$I_m$ and $K_m$ are the $m$-order modified Bessel functions, 
and $\epsilon=\epsilon_0\epsilon_r$ is the permittivity of the dielectric.  
The dipole potential is
\begin{equation}
V_{dip}(\rho,\phi,z)=\frac{p/d}{\pi\epsilon L}
\sum_{n=1}^{\infty} \sin\Bigr( \frac{n\pi z}{L}\Bigr) f_n(\rho,\phi;\theta_0,z_0),
\label{3d_1}
\end{equation}
where 
$f_n(\rho,\phi;\theta_0,z_0)=K_0(n\pi\rho/L)\sin(n\pi z_0/L) 
- K_0(n\pi|\bm{\rho}-\bm{\rho_0}|/L)\sin[n\pi(z_0+d\cos \theta_0)/L]$.
Here $|\bm{\rho}-\bm{\rho_0}|=(\rho^2+d^2\sin^2\theta_0 - 
2 \rho d \sin \theta_0 \cos \phi)^{1/2}$. 
\begin{figure}
\includegraphics[height=4.5cm,width=7.0cm]{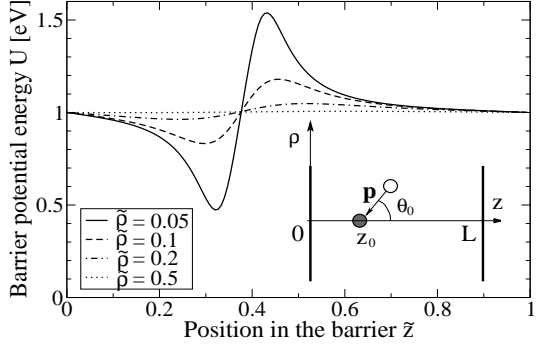}
\caption{\label{fig1} The potential energy barrier $U$ 
versus $\tilde{z}$ for different values of $\tilde{\rho}$. 
The dipole lies along the $z$--axis ($\tilde{z}_0=0.334$, $\theta_0 = 0$, and 
$\tilde{d}=0.0867$). We have used $p=|{\bf p}|=3.7$ D and $U_0=1$ eV
\cite{Dorneles2003}. The inset shows the geometry and the parameters 
of the model.}
\end{figure}
In Fig.~\ref{fig1}, the barrier potential $U$ is plotted versus $\tilde{z}=z/L$ 
for different values of $\tilde{\rho}=\rho/L$. The distortion of $U$ 
decreases as the radial distance $\tilde{\rho}$ increases, 
eventually becoming negligible when $\tilde{\rho}\simeq 1$. 
When the dipole flips 180$^o$, $U$ switches from
$\left(U_0 - eV_{dip}\right)$ 
to $\left(U_0 + eV_{dip}\right)$.

Next we follow \cite{Schmidlin1966,Scalapino1967} and use the WKB 
approximation to calculate ${\cal T}^{\pm}_{LR}$ 
corresponding to the two orientations of the dipole:
\begin{equation}
|{\cal T}_{LR}^{\pm}(\rho,\phi)|^2 \sim \exp\Bigr\{-2\int_{0}^{L} \mbox{d}z 
\sqrt{\frac{2m}{\hbar^2}\left(\Phi \mp eV_{dip}(\rho,\phi,z)\right)} \Bigr\},
\label{eq:sqrt}
\end{equation}
where $\Phi=U_0-\varepsilon_{k_{z}}$ and
$\varepsilon_{k_{z}}$ is the energy of the electron incident along 
the $z$ axis. 
We assume that $eV_{dip}\ll \Phi$ which allows us to expand 
the square root in the exponent in powers of $eV_{dip}/\Phi$, 
and to use the WKB approximation
which is valid for a potential that varies slowly along the tunneling 
direction. To a good approximation $\Phi$ is a constant 
representing the maximum barrier height \cite{Scalapino1967}. 
We average $|{\cal T}_{LR}^{\pm}|^2$ over the junction area $A$,
and we find to lowest order in $\left(eV_{dip}/\Phi\right)$: 
\begin{align}
\frac{\langle|{\cal T}_{LR}^{\pm}|^2\rangle}{|{\cal T}_{LR}^{0}|^2} &= \frac{1}{A}
\int_{0}^{2\pi}\mbox{d}\phi \int_{\rho_{min}}^{\rho_{max}}\mbox{d}\rho~\rho 
\exp\Bigr\{\pm  e\sqrt{\frac{2m}{\hbar^2 \Phi}} {\cal F}(\rho,\phi)\Bigr\}
\label{jc}
\end{align}
where $|{\cal T}_{LR}^{0}|^2\sim\exp(-2L\sqrt{2m\Phi/\hbar^2})$ is the 
square of the tunneling matrix element in the absence of impurities and
${\cal F}(\rho,\phi)=\int_{0}^{L} \mbox{d}z V_{dip}(\rho,\phi,z)$. 
${\cal F}(\rho,\phi)$ will be small due to the oscillation in $V_{dip}$, 
but it will be nonzero if there is asymmetry in the 
dipole's position along the $z$-axis or in its orientation $\theta_0$.
In the exponent ${\cal F}(\rho,\phi)$ can still have a noticeable 
effect on the noise. Integrating over $z$ yields
\begin{equation}
\frac{\langle|{\cal T}_{LR}^{\pm}|^2\rangle}{|{\cal T}_{LR}^{0}|^2}=\frac{1}{A}~
\int_{0}^{2\pi}\mbox{d}\phi \int_{\rho_{min}}^{\rho_{max}}\mbox{d}\rho~\rho 
\exp[\pm\beta~ W(\rho,\phi;\theta_0,z_0)],
\label{final_jc}
\end{equation}
where the constant 
$\beta=(p/d)/(\pi^2\epsilon)\times e\sqrt{2m/(\hbar^2 \Phi)}$ 
and 
$W(\rho,\phi;\theta_0,z_0)=
\sum_{n=1}^{\infty}[1-(-1)^n]f_n(\rho,\phi;\theta_0,z_0)/n$. 

The critical current $I_c$ is proportional to $\langle|{\cal T}_{LR}|^2\rangle$
\cite{AB1963}. Hence $I_c^{\pm}=I_c\langle|{\cal T}_{LR}^{\pm}|^2\rangle/
|{\cal T}_{LR}^{0}|^2$ where $I_c$ is the critical current in the absence of
any dipoles. The critical current fluctuations, defined as
$\Delta I_c(\theta_0,z_0)=I_c^+(\theta_0,z_0)-I_c^-(\theta_0,z_0)$,
are given by
\begin{align}
\frac{\Delta I_c}{I_c}&=\frac{L^2}{A}
\int_{0}^{2\pi}\mbox{d}\phi \int_{\tilde{\rho}_{min}}^{\tilde{\rho}_{max}}
\mbox{d}\tilde{\rho} ~\tilde{\rho} \{\exp[\beta~ 
W(\tilde{\rho},\phi;\theta_0,\tilde{z}_0)]\nonumber \\
&-\exp[-\beta~ W(\tilde{\rho},\phi;\theta_0,\tilde{z}_0)]\}
\equiv \frac{L^2 \Delta g(\theta_0,\tilde{z}_0)}{A},
\label{Ic_fluct}
\end{align}
where the $L^2$ factor comes from introducing 
$\tilde{\rho}=\rho/L$. 

To find the critical current noise power $S_{I_c}$, we 
assume that each dipole produces a Lorentzian spectrum \cite{Kogan96}:
\begin{equation}
\frac{S^{(i)}_{I_c}(f)}{I_c^2}=\left\langle\left(\frac{\Delta I_c}{I_c}\right)^2
\right\rangle \frac{4P_{+}P_{-}T_{1}}{1+\omega^2T_1^2},
\end{equation}
where $i$ denotes the $i$th dipole, $P_{\pm}=\exp\left(\mp E/2k_BT\right)/Z$ is the
Boltzmann probability of being in the upper (lower) state of the TLS, 
the partition function $Z=2\cosh\left(E/2k_BT\right)$, and
$T$ is temperature. 
We average over the distribution of TLS to find the low frequency 
($\omega\tau_{min}\ll 1$) $1/f$ noise power: 
\begin{eqnarray}
\frac{S_{I_c}(f)}{I_c^2}
&\simeq & \int_{0}^{E_{M}}\mbox{d}E \int_{0}^{\infty} \mbox{d}T_1
~\frac{P_0 V}{2T_1}\frac{\langle(\Delta I_c/I_c)^2\rangle}
{\cosh^2\left(\frac{E}{2k_B T}\right)} \frac{T_1}{1+\omega^2T_1^2}\nonumber\\
&\simeq& \frac{P_0 k_B T}{4f} \frac{L^5}{A}\langle \Delta g^2 \rangle.
\label{SIc}
\end{eqnarray}
where $V=AL$, $E_{M}\gg k_BT$, and we can neglect the
factor of $1/\sqrt{1-\tau_{min}(E)/T_1}$ because $\tau_{min}/T_1\ll 1$
in the region ($E/(2k_BT)>0.1$, $\omega T_1>0.1$) that dominates the integral.
$\langle \Delta g^2\rangle$ contains the average over each dipole's orientation
and position along the $z-$axis:
$\langle \Delta g^2\rangle = 1/\left[2(\tilde{z}_{0_{M}}-\tilde{z}_{0_{m}})\right] 
\times\int_{0}^{\pi}\mbox{d}\theta_0 \sin \theta_0 \int_{\tilde{z}_{0_{m}}}^
{\tilde{z}_{0_{M}}}\mbox{d}\tilde{z}_0~\Delta g^2(\theta_0,\tilde{z}_0)$,
where $\tilde{z_0}$ lies between $\tilde{z}_{0_{m}}=\tilde{d}$ and 
$\tilde{z}_{0_{M}}=1-\tilde{d}$ to ensure that the dipole lies entirely in 
the dielectric region. Note that the critical current noise power scales 
as $L^5/A$. Although the $A^{-1}$ dependence is well known experimentally
\cite{Harlingen2004,Savo1987,Wellstood,Wellstood2004}, it would interesting 
to check experimentally the $L^5$ scaling of the noise predicted by our model. 

\begin{figure}
\includegraphics[height=4.5cm,width=7.0cm]{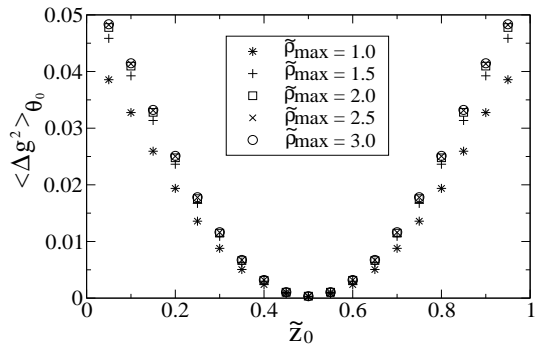}
\caption{\label{fig2} 
$\Delta g^2$ averaged over the dipole orientation
angle $\theta_0$ versus the dipole position $\tilde{z}_0$.} 
\end{figure}

To estimate $\langle\Delta g^2\rangle$, we evaluated 
the integrals numerically with 
$\tilde{\rho}_{min}=0.1$, which is comparable to an atomic 
radius. As Fig.~\ref{fig1} shows, 
the lower cutoff $\tilde{\rho}_{min}$ is needed
because the effect of the dipole on the tunnel barrier 
is no longer weak for $\tilde{\rho}\stackrel{<}{\sim} 0.1$. 
Fig.~\ref{fig2} shows $\Delta g^2$ averaged over $\theta_0$ versus position
for various values of $\tilde{\rho}_{max}$. The results
overlap for $\tilde{\rho}_{max} > 2$. Note 
that the largest contribution to $S_{I_c}$ comes 
from dipoles near the electrodes.

Our numerical estimates follow. We obtain 
$\langle\Delta g^2\rangle=1.5782\times 10^{-2}$ 
for $p=3.7$ Debye (which corresponds to the dipole moment of 
an OH$^-$ impurity \cite{Golding1979}), $d=0.13$ nm, 
$\epsilon_r=10$, $\tilde{\rho}_{max}=4.0$, and $U_0=1$ eV \cite{Dorneles2003}. 
For a Josephson junction 
with $A=1\;\mu$m$^2$, $L=1.5$ nm, $P_0=10^{45}$ (Jm$^3$)$^{-1}$ 
\cite{Phillips}, $f=1$ Hz, and $T=100$ mK, we obtain a noise 
power of $S_{I_c}/I_c^2=4.13\times 10^{-14}$ Hz$^{-1}$, in good 
agreement with the recent tunnel junction resistance measurements 
by the Delft group \cite{Mooij2006} who deduced a value of 
$2.0\times 10^{-14}$ Hz$^{-1}$ for the normalized critical current 
noise power using similar junction parameters. They also found 
that the noise varied linearly with temperature in agreement with
Eq.~(\ref{SIc}). Other measurements 
in large superconducting junctions find larger $S_{I_c}$
than those of the Delft group \cite{Savo1987,Wellstood,Wellstood2004}. 
Using Eq.~(\ref{SIc}), we find the critical current noise in larger 
junctions at 4.2 K to be roughly 100 times lower than the experimental 
values in \cite{Savo1987,Wellstood2004}. The noise level difference 
between the Delft group \cite{Mooij2006} and those working 
with larger junctions may be due to differences in fabrication 
techniques. For example, our calculation indicates that if the 
effective thickness $L$ of the tunnel barrier in the Nb-AlO$_x$-Nb 
junctions \cite{Savo1987,Wellstood} or Nb-NbO$_x$-PbIn junctions 
\cite{Wellstood,Wellstood2004} were larger by only a factor of 2 
with respect to our $L=1.5$ nm, this ``discrepancy'' 
would be essentially removed. A 
$T^2$ dependence was observed in Ref.~\cite{Wellstood2004} which 
may imply an additional mechanism for the noise \cite{Faoro2006b}. 

The dipoles that produce critical current noise are also responsible
for the fluctuations in the induced charges on the superconducting 
electrodes that produce charge noise. The dipole making 180$^o$ flips 
induces charge fluctuations ($\Delta Q=|2 p \cos\theta_0/L|$) on the 
electrodes. The time series of the charge is a two-state random telegraph 
signal with a transition rate $T_1^{-1}$ resulting in a Lorentzian charge 
noise spectrum \cite{Kogan96}. The charge noise results from 
averaging over the TLS density of states \cite{Kogan96,Faoro2006}. 
At low frequencies one obtains 
\begin{equation}
\frac{S_Q(f)}{e^2}=\frac{1}{3}V P_0 k_BT \Bigr(\frac{p}{eL}\Bigr)^2\frac{1}{f}.
\label{eq:SQ_LF}
\end{equation}
We estimate $S_Q/e^2=1.83\times10^{-3}$ Hz$^{-1}$, in good agreement with 
the experimental values ranging from $10^{-4}$ Hz$^{-1}$ \cite{Martinis1992} 
to approximately $4\times 10^{-4}$ Hz$^{-1}$ \cite{Zorin1996}. 
Using Eqs.~(\ref{eq:SQ_LF}) and (\ref{SIc}), we obtain
the ratio between $S_Q/e^2$ and $S_{I_c}/I_c^2$ at low frequencies:
\begin{equation}
\frac{S_Q/e^2}{S_{I_c}/I_c^2}=B\biggl( \frac{pA}{eL^3}\biggr)^2,
\label{ratio}
\end{equation}
where $B=4/(3\langle \Delta g^2\rangle)$. 
For the physical parameters listed above, we obtain $B\simeq 84.5$ 
and $(S_Q/e^2)/(S_{I_c}/I_c^2)\simeq 4.4\times 10^{10}$. This is consistent
with the value of $\simeq 2\times 10^{10}$ deduced from experimental 
measurements of $S_Q$ \cite{Zorin1996} and $S_{I_c}$ \cite{Mooij2006}, 
though the $S_Q$ and $S_{I_c}$ measurements were not made on the same samples. 

\begin{figure}
\includegraphics[height=5cm,width=7.5cm]{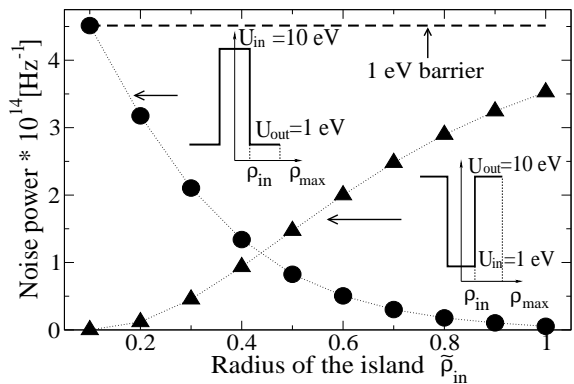}
\caption{\label{fig3} 
Normalized critical current noise for a nonuniform tunnel barrier with 
two islands ($U_{in}=10$ eV and $U_{out}=1$ eV and vice-versa) versus the radius 
of the inner island $\tilde{\rho}_{in}$. The dashed line corresponds to noise
from a dipole in a square barrier with a height of 1 eV. The critical current
fluctuations in all cases are normalized by the critical current through
a uniform square barrier of 1 eV with no dipole present.} 
\end{figure}

Our calculation of the critical current noise assumed that in the
absence of the dipoles, the tunnel barrier is uniform. 
However, local fluctuations in oxide thickness or barrier height 
make the tunnel barrier nonuniform and may result in ``pinholes''
in the barrier \cite{Dorneles2003}. 
We can investigate the sensitivity of the critical current noise 
to the nonuniformity of the tunnel barrier by placing the fluctuating 
dipole in a cylindrical island of radius $\rho_{in}$
with the axis of the cylinder along $\hat{z}$. We set the tunneling 
barrier of the island to be different from that outside the island: 
$U(\rho)=U_{in}$ for $0 \leq \rho <\rho_{in}$ and $U(\rho)=U_{out}$ 
for $\rho_{in}\leq \rho \leq \rho_{max}$.  We can model a pinhole by 
having $U_{in}<U_{out}$. The results in Fig.~\ref{fig3}
show that the noise is enhanced as the radius $\tilde{\rho}_{in}$
of the inner cylinder grows because the dipole distortions of the barrier
potential are a larger fraction of $U_{in}$ than of $U_{out}$.
Unlike before, we do not expand the argument of the square root 
in Eq.~(\ref{eq:sqrt}). 
When $\tilde{\rho}_{in} \rightarrow \tilde{\rho}_{max}$, we obtain 
our previous result for a dipole in a square 1 eV barrier.

In the other case where $U_{in} > U_{out}$, and the dipole is on the inner island,
then the noise decreases as $\tilde{\rho}_{in}$
increases. This is shown in Fig.~\ref{fig3} with $U_{in}=10$ eV and
$U_{out}=1$ eV. The large barrier on the inner island reduces the amount
of tunneling current through the island and hence limits the noise
due to the fluctuating two level system. In the limit
($\tilde{\rho}_{in} \rightarrow \tilde{\rho}_{min}$) that the inner
island disappears, we recover our previous result for a dipole in a square
barrier of 1 eV. 
This investigation shows that the noise power can be affected by the
nonuniformity of the barrier, and could explain why several experimental 
groups have measured different values of $S_{I_c}/I_c^2$ 
\cite{Wellstood2004,Harlingen2004,Mooij2006}. 

To conclude, we have used a microscopic model of fluctuating
dipoles in the tunnel barrier to calculate the critical current 
noise in Josephson junctions. By considering a simple tunnel barrier 
potential with a fluctuating electric dipole associated with 
the presence of two-level systems in the barrier, we estimated the
critical current noise which was in good agreement with recent 
experiments. We also found that nonuniformities such as
pinholes in the barrier can affect the noise. It would be 
interesting to test our prediction that the critical current 
noise goes as $L^5$, and to see if the ratio of the charge 
noise to critical current noise for the same junction
is given by Eq.~(\ref{ratio}) at low frequencies. 

This work was supported by ARDA through ARO Grant W911NF-04-1-0204,
and by DOE grant DE-FG02-04ER46107. 
We wish to thank John Martinis, Alex Maradudin, and Fred 
Wellstood for useful suggestions.

\end{document}